\documentclass[preprints,article,accept,pdftex,moreauthors]{Definitions/mdpi} 

\usepackage{microtype,siunitx,booktabs}
\usepackage{amssymb}
\usepackage{mathabx}
\usepackage{amsmath}

\newcommand{\beq}{\begin{equation}}
\newcommand{\eeq}{\end{equation}}
\newcommand{\bea}{\begin{eqnarray}}
\newcommand{\eea}{\end{eqnarray}}
\newcommand{\mwater}{M_{\text{H}_2\text{O}}}
\newcommand{\fwater}{f_{\text{H}_2\text{O}}}

\firstpage{1} 
\pubvolume{1}
\issuenum{1}
\articlenumber{0}
\pubyear{2023}
\copyrightyear{2022}
\externaleditor{{Academic Editor:} Firstname Lastname}
\datereceived{} 
\dateaccepted{} 
\datepublished{} 
\hreflink{https://doi.org/} 
\pdfoutput=1

\Title{Multiverse Predictions for Habitability: Planetary~Characteristics}

\TitleCitation{Multiverse Predictions for Habitability: Planetary Characteristics}

\Author{McCullen Sandora $^{1,}$*,
Vladimir Airapetian $^{2,3}$,
Luke Barnes $^{4}$ and
Geraint F. Lewis $^{5}$} 

\AuthorNames{McCullen Sandora,
Vladimir Airapetian,
Luke Barnes and
Geraint F. Lewis}

\AuthorCitation{Sandora, M.; Airapetian, V.; Barnes, L.; Lewis, G.F.}

\address{
$^{1}$ \quad Blue Marble Space Institute of Science, Seattle, WA 98154, USA\\
$^{2}$ \quad Sellers Exoplanetary Environments Collaboration, NASA Goddard Space Flight Center, Greenbelt, MD, 20771, USA\\
$^{3}$ \quad Department of Physics, American University, Washington, DC, USA\\
$^{4}$ \quad School of Science, Western Sydney University, Locked Bag 1797, \mbox{Penrith South DC, NSW 2751, Australia}\\
$^{5}$ \quad Sydney Institute for Astronomy, School of Physics, A28, The University of Sydney, NSW 2006, Australia}

\corres{Correspondence: mccullen@bmsis.org}

\abstract{Recent detections of potentially habitable exoplanets around sunlike stars demand increased exploration of the physical conditions that can sustain life, by whatever methods available.  Insight into these conditions can be gained by considering the multiverse hypothesis; in a multiverse setting, the probability of living in our universe depends on assumptions made about the factors affecting habitability.  Various proposed habitability criteria can be systematically considered to rate each on the basis of their compatibility with the multiverse, generating predictions which can both guide expectations for life's occurrence and test the multiverse hypothesis. Here, we evaluate several aspects of planetary habitability, and show that the multiverse does indeed induce strong preferences among them.  We find that the notion that a large moon is necessary for habitability is untenable in the multiverse scenario, as in the majority of parameter space, moons are not necessary to maintain stable obliquity.  Further, we consider various proposed mechanisms for water delivery to the early Earth, including delivery from asteroids, both during giant planet formation and a grand tack, delivery from comets, and oxidation of a primary atmosphere by a magma ocean.  We find that, depending on assumptions for how habitability depends on water content, some of these proposed mechanisms are disfavored in the multiverse scenario by Bayes factors of up to several hundred.}


\begin{document}

\section{Introduction}
	
	This is the sixth entry in a series of papers aimed at using the multiverse framework suggested by modern cosmology to deduce testable predictions.  Since its inception within the realm of modern science, the multiverse hypothesis, i.e., the idea that the laws of physics may be different elsewhere, has attracted considerable controversy.  Most of this stems from the fact that in all likelihood, we are confined to this universe, with no prospect of observing another universe directly, precluding any possibility of directly verifying or falsifying this theory.  However, this situation is far from hopeless, as indirect evidence can be garnered to either support or refute the theory.
	
	Our essential reasoning is the following: the multiverse framework must be able to account for our observations, which consist of measurements of our values of the fundamental constants of nature.  In the multiverse, the probability of measuring a particular value is directly proportional to the number of universes created with that value, multiplied by the number of observers that arise in that universe.  Most of the predictive power lies in this latter factor, as the estimate of the expected number of observers in universes with differing fundamental constants hinges drastically on assumptions we make about what the requirements for habitability are.  There are no shortage of proposed environmental factors purported to be essential for habitability, taken here to mean the emergence and continued sustenance of primitive and complex life \citep{cockell2016habitability,airapetian2018terrestrial,airapetian2020impact,schwieterman2018exoplanet,schwieterman2019limited,rimmer2018origin}.  The major uncertainty is in quantifying these factors for our universe accurately enough to make reliable extrapolations for their contribution in other universes.  Thus, we necessarily rely on many approximate encapsulations.  
	
	Our strategy is then to systematically incorporate as many proposed habitability criteria into this framework to determine which are compatible with the multiverse, and which are not.  For this, we have started with crude measures of habitability, and are gradually accruing more sophisticated requirements.  The notions we consider in this paper can generically be categorized as relating to planetary habitability.  This is a direct extension of our previous work \citep{mc2}, which already considered many basic considerations.  There, we delineated the requirements for the production of planets in nature, and concluded that planets are ubiquitous features of the multiverse.  We then asked whether the probabilities of observing our constants depend greatly on assumptions about which planetary size and temperature are habitable, and found that they do not.
	
	In this paper, we extend our approach to more nuanced aspects of planetary habitability.  We begin by discussing planetary orbital elements; we consider eccentricity in Section \ref{Eccentricity} and obliquity in Section \ref{Obliquity}, specify their habitable ranges according to several criteria, and determine how the distribution of these varies with fundamental constants.  Along these lines, we consider the effect a large moon has on the axial stability of a planet, the conditions for obliquity to be unstable without one, and the range of parameters for which large moons can be formed and retained.  We consider one popular habitability hypothesis, that a large moon is essential for complex life, on account of it stabilizing planetary obliquities.  We find that this does not make much sense in a multiverse context, because for the majority of parameter space of values for the physical constants, planet obliquities are already stable without a moon.  In these universes, every planet would meet these qualifications to be potentially habitable, whereas, according to the large moon hypothesis, in our universe, only a small fraction do.  This places a strong penalty on our universe relative to others, and begs the question of why we would find ourselves in one of the few less fertile universes, amongst a majority that do not suffer this affliction.
	
	We next discuss the water content of terrestrial planets in Section \ref{Water}.  There are two aspects of this that need to be considered: the first is the source of water in the inner planetary system, which is not definitively known at present.  The second is how habitability depends on the amount of water on a planet.  We consider various hypotheses for both of these aspects, and what they suggest for the distribution of life throughout the multiverse.  The delivery mechanisms considered here include asteroids, both during giant planet formation or a grand tack, via comets, and oxidation of a primary atmosphere during a magma ocean phase.  The fact that Earth's water content is just enough to cover a reasonable fraction of the planet's surface area is a special arrangement, and has been explained through selection effects, either as the arrangement that maximizes bio-productivity, or because it is situated near a habitability threshold with a strong preference for either much wetter or drier planets.  We consider six proposed dependences of habitability on surface water fraction, and find that several combinations of these with the proposed delivery mechanisms are incompatible with the multiverse.  
		
	Our efforts here add several predictions for habitability resulting from the multiverse framework.  First, according to the multiverse, stable obliquity is not required for complex life.  Secondly, several proposed mechanisms for water delivery, and how habitability depends on water content, are incompatible with the multiverse, though which depend on assumptions made for other characteristics, as detailed below.
		
	The main observables we are concerned with are the probabilities of measuring our values of the five constants that have the most bearing on our mesoscopic world: the fine structure constant $\alpha$ governing the strength of electromagnetic interactions, and four mass ratios: the electron to proton mass ratio $\beta=m_e/m_p$, the proton to Planck mass ratio $\gamma=m_p/M_{pl}$, and the ratios of the up and down quark masses to proton mass $\delta_u=m_u/m_p$, $\delta_d=m_d/m_p$.  Additionally, $\kappa=1.1\times10^{-16}$ parameterizes galactic density, though we do not consider its variation in this paper.  {The probability of measuring the observed values as a function of habitability condition} $\mathbb H$ {is taken as} $\mathbb P(x_\text{obs}|\mathbb H)=\min(P(x<x_\text{obs}|\mathbb H),P(x>x_\text{obs}|\mathbb H))$, {with probability density function} $p(x)\propto p_\text{prior}(x)\,\mathbb H(x)$.  {We take the prior to be log-uniform in the mass ratio variables, based on theoretical arguments that this is a natural consequence of the strong coupling that sets the proton mass} \citep{schellekens} {and renormalization group flow which sets the other masses} \citep{leptonland}.  {We use a uniform distribution for the force strength} $\alpha$, {yielding} $p_\text{prior}\propto 1/(\beta\gamma\delta_u\delta_d)$.  {The influence of the prior on our analyses is investigated in an upcoming publication} \citep{mc8}.

    {Here the habitability of a universe is defined as the number of observers a universe produces over the course of its evolution, and can be decomposed in a Drake-style equation as} $\mathbb H = N_\text{obs}= N_\star\,n_e\,f_\text{life}\,f_\text{int}\,N_\text{int}$, {where} $N_\star$ {is the number of habitable stars per universe,} $n_e$ {is the average number of habitable planets per habitable star,} $f_\text{life}$ {is the fraction of habitable planets that develop life,} $f_\text{int}$ {is the fraction of biospheres that develop intelligent observers, and} $N_\text{int}$ is the average number of observers that arise on planets that develop intelligence.  For most of this paper, we take as a baseline the habitability criteria that were found to produce the best account of our observations in our previous analyses.  These are: the {\bf yellow} {condition, that a relatively narrow energetic band is required for photosynthesis}~\citep{mc1}, {which restricts the mass range of stars to be counted in} $N_\star$.  The {\bf entropy} {condition, that the habitability of a planet is proportional to the entropy it receives from starlight}~\citep{mc3}, {so that} $f_\text{life}\propto \Delta S$.  {The} {\bf C/O} {condition, that a carbon to oxygen ratio of close to 1 is required}~\citep{mc5}.  {Furthermore, in this paper we employ the {\bf temp} and {\bf terr} {conditions, which restrict our analysis to planets that are temperate (capable of retaining liquid surface water) and terrestrial (having small secondary, high molecular mean atmospheres), respectively, which further restrict our definition of} $n_e$ \citep{mc2}.  Though these last two conditions were not found to improve agreement with observations, they significantly simplify our analysis, and will be relaxed in a future publication.  Collectively, we refer to the set of these habitability factors as the {\bf yellow + entropy + terr + temp + C/O} condition.  We parameterize stellar mass in terms of Chandrasekhar mass as $\lambda=M_\star/M_\text{Ch}$.  In addition, we employ the astronomical notation for characteristics of solar system bodies, or their analogues in other stellar systems: The sun (star) $\Sun$, Earth (temperate, terrestrial planet under consideration) $\Earth$, moon $\Moon$, Jupiter (innermost gas giant planet) $\Jupiter$, and Saturn (second gas giant) $\Saturn$.  
	
	\section{Eccentricity}\label{Eccentricity}
	
	Orbital eccentricity, $e$, is the first aspect of planetary habitability we consider here. This factor measures the deviation from circularity of a planet's orbit ($e=0$ being perfectly circular).  It is reasonable to expect that there is some value of eccentricity beyond which a planet's habitability drastically suffers, either through the effects an annual change in instellation would have on the atmosphere, or through interactions with neighboring planets.  Typical planetary eccentricities are a byproduct of the planet formation process, with some spread around the average value.  If the outcome of this process lead to significantly higher eccentricities, as can be the case for different values of the physical constants, then a substantially greater fraction of planets would have been precluded from long-term habitability.  Here, we investigate the strength of this effect, and assess its potential contribution in selection effects of our observed constants.
	
	Earth's current eccentricity is $e=0.0167$, though it reached values of 0.07 prehistorically~\citep{berger1976obliquity}.  This latter value serves as a maximally conservative potential upper bound on habitable eccentricities, since we know Earth has maintained continuous habitability over the course of its lifetime, though most proposed upper bounds are considerably larger than this.  To determine the fraction of planets below a given eccentricity, we first need the form of the eccentricity distribution.  This is actually a contentious point in planetary science, with a number of proposed distributions of varying degree of justification pervading the literature \citep{winn2015occurrence,kipping2013parametrizing}.  The distribution we will use was derived in \citep{tremaine2015statistical} using a statistical mechanical treatment of planetary orbits.  This yields $p(e)=4 e/\bar e^2 \exp(-2 e/\bar e)$, and was shown to be a good fit to the eccentricity distribution that arises from simulations.  This distribution gives a fraction of planets below a certain maximum eccentricity as
	\beq
	f(e_\text{max}) = 1-\left(1+\frac{2\,e_\text{max}}{\bar e}\right)\exp\left(-\frac{2\, e_\text{max}}{\bar e}\right)
	\eeq
	This tends to 1 for $\bar e \ll e_\text{max}$ and $2e_\text{max}^2/\bar e^2$ for $\bar e \gg e_\text{max}$.  Of particular note is the fact that the tail of this distribution is considerably fatter than the ever popular Rayleigh distribution, though the use of this distribution instead has little effect on our results.
	
	Next is to determine the quantities $\bar e$ and $e_\text{max}$ in this expression, and their dependence on physical constants.  The mean eccentricity $\bar e$ is dictated by planet formation in terms of the Hill eccentricity \citep{greenzweig1990accretion}; this accounts for the usual thermal spread of planetesimal velocities compared to the Keplerian velocity, and scales as $\bar e\sim R_\text{Hill}/a\sim (M_\text{planet}/M_\star)^{1/3}$, where $R_\text{Hill}$ is the Hill radius (the outer limit of stable bound orbits) and $a$ is the semimajor axis.  To relate mean eccentricity to physical constants, it remains to find a suitable expression for planet mass.
	
	An expression for the typical planet mass that results from a giant impact phase was derived in \citep{schlichting2014formation}.  This is
	\beq
	M_\text{planet}\sim\left(\frac{4\,\pi\,\Sigma\,a^{5/2}\rho^{1/6}}{M_\star^{1/2}}\right)^{3/2}
	\eeq
	Here $\rho$ is the density of solid material and $\Sigma$ is the protoplanetary disk density.  In our previous works, these quantities were expressed in terms of fundamental constants, allowing us to infer how their properties change in other universes.  For convenience, we have compiled all previously discussed quantities relevant for our discussion in Table \ref{tablefund} in the appendix.  Applying these relations, and specializing to planets within the habitable zone, where liquid surface water is possible, we find the average eccentricity to be given by
	\beq
	\bar e =159.4\,\frac{\kappa^{1/2}\,\lambda^{17/16}\,\gamma^{3/8}}{\alpha^{7/2}\,\beta^{5/4}}\label{bare}
	\eeq
	The prefactor has been fixed to reproduce the observed average eccentricity of 0.04, which is both the solar system and observed galactic average \citep{van2019orbital}.  Bear in mind that this represents the primordial eccentricity distribution; it is presumed that subsequent interactions, such as gas drag \citep{kominami2002effect}, may alter the eccentricity somewhat, but not to the extent that an additional scale is introduced.
	
	This must now be compared to the maximum eccentricity compatible with habitability.  This maximum value will surely depend on the details of each planetary system, but treating it as a fixed number will suffice to give the expected fraction of systems with habitable eccentricities.  The threshold can come from two effects: orbital stability, and atmospheric stability.  We detail each of these in turn.
	
	The simplest approximate orbital stability criterion is known as Hill stability.  Strictly applicable to three body systems, it delineates that orbits are stable if the two planets never pass within their mutual Hill radii, which would impart significant angular momentum change between the two planets.  This condition was found to give the following restriction on eccentricities: $e\lesssim (M_\text{planet}/M_\star)^{1/3}$ \citep{marchal1982hill}.  Note that this scales in the same manner as the average eccentricity; therefore, the fraction of planets which are Hill stable is independent of physical constants.  We will discuss more sophisticated stability determination criteria in the next section, but note here that none substantially alter our findings.
	
	The second effect to consider is stability of the planetary atmosphere.  Eccentricity causes an annual difference in instellation, which if extreme enough can cause a planet to periodically exit the habitable zone.  On Earth, temperature differences due to eccentricity are far dwarfed by those induced by obliquity, but eccentricity perturbations from Milankovitch cycles lead to noticeable effects on Earth's past climate \citep{hays1976variations}.  This arises because nonlinear Earth system responses make a planet more sensitive than one would naively expect~\citep{imbrie1993structure}.  Eccentricity also shifts the annually averaged stellar flux a planet receives, but this effect can be nullified by shifting our attention to planets with semimajor axes that appropriately compensate for this.  To estimate the annual temperature change, note that the difference between a planet's  perihelion and aphelion is given by $\Delta a = 2 e a$, and that the temperature of a planet scales as $T\propto 1/\sqrt{a}$.  The annual fractional difference in temperature is then given by $\Delta T/T \approx e$ (valid for small $e$; using a more generally valid expression has little effect).  This disregards the thermal inertia of the planetary system, which will temper the response and is known to be important \citep{dressing2010habitable}, and so this can be viewed as a pessimistic estimate of the effect.
	
	The allowable annual fractional temperature difference can be estimated by analogy to the boundaries of the habitable zone; it was found in \citep{leconte95} that inside 0.95 AU, a runaway greenhouse effect is induced, and outside 1.67 AU, a runaway icehouse effect occurs \citep{kastHZ}.  Translating into temperature, this corresponds to $\Delta T/T\approx e_\text{max}\approx 0.25$.  Again, this is a pessimistic estimate, as here the runaway effects will only be active for part of the orbit, and may be compensated or reversed during the other part.  Indeed, climate modeling suggests that eccentricities may be significantly higher while maintaining surface liquid water, at least through a part of the year; Ref. \citep{palubski2020habitability} conclude that planets with eccentricities as high as 0.5 can still host liquid water, albeit with highly restrictive orbital parameters.
	
	If the primary requirement for habitability is maintenance of surface liquid water, then the range of allowable temperatures will also depend on atmospheric pressure (albeit logarithmically, as long as the pressure exceeds the triple point of water of 0.01 bar, below which liquid water cannot exist on a planet's surface).  As we do not take variations in atmosphere into account in this paper, for now we take $e_\text{max}$ to be a fixed quantity, independent of physical constants.  With this, we are ready to compute the effects placing a restriction on eccentricity have on the probability of measuring the observed values of our constants.  For reference, we first display the probabilities without including this criterion.  With the {\bf yellow + entropy + terr + temp + C/O} condition:
	\bea
	\mathbb{P}(\alpha_{obs})=0.24,\,\, 
	\mathbb{P}(\beta_{obs})=0.06,\,\, 
	\mathbb{P}(\gamma_{obs})=0.05,\,\,
	\mathbb{P}(\delta_{u\phantom{!}obs})=0.29,\,\,
	\mathbb{P}(\delta_{d\phantom{!}obs})=0.37\label{basep}
	\eea
	The effects of including a maximum habitable eccentricity are displayed in Figure~\ref{emax}.  As can be seen, the effects hardly register at all.  This is because, even though there are regions of parameter space where eccentricities are much larger, these are extreme, and/or disfavored for other reasons.  As such, no prediction can be made as to whether life on high eccentricity planets is possible or impossible.  In light of this, the simplifying assumptions we made in this section are shown to be adequate, as none are capable of substantially amplifying the importance of this effect.
		\begin{figure}[H]
			\includegraphics[width=0.95\textwidth]{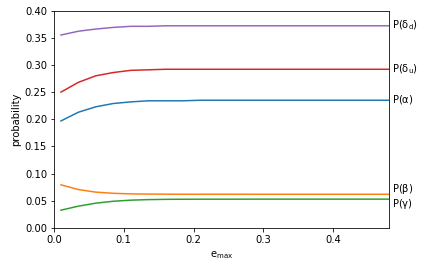}
			\caption{The effect of varying $e_\text{max}$ on the probabilities of observing our constants.  Above $e_\text{max}=0.1$, probabilities are changed by at most $4\%$.}
			\label{emax}
		\end{figure}

	\section{Obliquity}\label{Obliquity}
	
	It is commonly said that the Earth's moon may be partly responsibly for maintaining habitable conditions on Earth \citep{rareearth}.  The most commonly attributed reason given is that the moon stabilizes Earth's obliquity, though sometimes its effects on tidal forces \citep{balbus2014dynamical}, or Earth's geodynamo \citep{andrault2016deep} are invoked.  Here, we focus on the first explanation.
	
	The overall influence obliquity has on a planet's habitability is uncertain and disputed to this day.  \citep{kang2019wetter} points out that water loss on high obliquity planets would be enhanced as a result of the larger annual temperature swings, potentially cutting any habitable phase short.  In contrast, \citep{dong2019role} find that obliquity has a relatively minor effect on atmospheric escape.  \citep{colose2019enhanced} argue that high obliquity planets may even have increased habitability, as their higher temperatures increase the width of the habitable zone.  Therefore, we may contemplate the implications of this controversial habitability criterion for the probability of observing our constants, in the hope of contributing to this debate from a different perspective.
	
	In this section, we discuss the following aspects of obliquity stability, and how these depend on fundamental constants: the magnitude and timescale of obliquity variations, the properties of moons, their potential for obliquity stabilization, and their occurrence rates.  Moon occurrence rate is dictated by three effects: the presence of a giant impact phase, the allowance of relatively gentle collisions, and long retention timescales.  Then, we synthesize these considerations into a factor that restricts our calculations to planets with stable obliquity.  We find that this does not affect the probabilities of observing our values of the constants much. However, we find that the probability of being around such a large star becomes quite low, as planets around smaller stars do not require moons to have stable obliquities.  Additionally, we find that in the majority of parameter space, moonless planets have stable obliquities.  This places strain on explaining the orbital arrangement we observe, where only the Earth-moon system has stable obliquity, as the majority of observers in the multiverse perspective experience their bare planet to be stable irrespective of any moon.
	
	\subsection{Why Is Obliquity So Unstable?}
	
	Earth's current obliquity is 23.4$^\degree$, and varies by $2.4^\degree$ on 40 kyr timescales \citep{berger1976obliquity}.  Though slight, this causes significant effects on climate, leading to measurable changes in ice cover and global temperature \citep{hays1976variations}.  As pointed out in \citep{laskar1993stabilization}, the presence of the moon serves as a major stabilizing factor for Earth's obliquity, which otherwise would swing with much larger amplitude.  For instance, Mars, without a large moon, experiences obliquity swings that are 10 times larger \citep{ward1973large}.  
	
	Such volatility need not have been the case, however.  It occurs only because the timescale of obliquity variations overlaps with secular resonances of the outer solar system planets; this causes a semi-coherent accumulation of forces over the course of the inner planets' orbits that dramatically enhances the total variation.  The obliquities of the outer planets, in the absence of such perturbing resonances, are quite stable \citep{laskar1993chaotic}.  Since these two different frequencies depend on fundamental constants in different ways, they need not always overlap with each other.  If not, then the obliquities of the inner planets would automatically be stable, with or without a large moon.  If obliquity is truly important for the development of complex life, then these universes would host many more candidates for life bearing planets, particularly because the presence of a large moon is estimated to be so rare.
	
	The frequency of Earth's obliquity variation is given by the sum of two terms as \citep{laskar1993stabilization} 
	\beq
	\Omega_\text{ob} = k_\text{ob}\,\omega_\text{rot} \frac{R_\Earth^3}{M_\Earth}\left(\frac{M_\Sun}{a_\Earth^3}+\frac{M_\Moon}{a_\Moon^3}\right)
	\eeq
	Here $\omega_\text{rot}$ is the Earth's rotation frequency, $a_\Moon$ is the Earth-moon distance, and $k_\text{ob}\sim0.2$ is a constant that depends on rotational configuration.  The observed values are $\Omega_\Earth = 15.9''/$yr, $\Omega_\Moon = 34.6''/$yr, $\Omega_\text{ob}=50.4''/$yr, compared with the secular resonance frequency \mbox{$\Omega_\Jupiter=25.8''/$yr}.  Several features of this arrangement are quite striking.  Firstly, the magnitude of the two contributions are roughly equal, though their origins are very disparate.  Additionally, without the moon's contribution, the precession frequency would lie in the unstable regime, and the moon's additional increase lifts it out of the dangerous regime only by a factor of 2.  Lastly, as noted in \citep{waltham2004anthropic}, if the Earth were moonless but rotating twice as fast, instability would equally well have been avoided, and large moons can induce tides that cause their host planet to spin down faster.
	
	We can determine the dependence of each of these frequencies on physical constants, which will allow us to extrapolate their values to other universes, and determine in which cases obliquities are stable.  The first, $\Omega_\Earth$, is relatively straightforward;  the factor $R_\Earth^3/M_\Earth$ can be replaced by the expression for the density of rock $\rho\sim\alpha^3 m_e^3 m_p$ given in \citep{PL}.  The magnitude of the rotation frequency is set by the breakup speed, beyond which centrifugal forces cause the planet to spin apart.  Rocky bodies are generically expected to spin at slightly lower than this frequency as a result of formation processes \citep{li2020planetary}.  This is given by $\omega_\text{rot}\sim \sqrt{G\rho}$, independent of planet size to the extent that density is, in line with the observation that most rocky solar system bodies rotate with a period of about ten hours~\citep{warner2009asteroid}.  
	
	With these expressions, the precession frequency of a moonless rocky planet in the habitable zone is found to be
	\beq
	\Omega_\Earth = 4.8\times10^{-4}\,\frac{\alpha^{27/2}\,m_e^{9/2}\,M_{pl}^{1/2}}{\lambda^{17/4}\,m_p^4}
	\eeq
	
	The reason the obliquities of inner planets are unstable is because the precession frequencies overlap with solar system secular frequencies.  These resonances are quite broad \citep{laskar1993chaotic} and extend from $0''/$yr to a maximum value, so that as long as the frequency scales overlap, instability can be expected to occur.  This maximum secular frequency is set by the giant planet which is largest and nearest to the star, i.e., Jupiter or its analogue.  The secular frequency is given by \citep{murray1999solar}:
	\beq
	\Omega_\Jupiter = k_\Jupiter \frac{M_\Jupiter}{M_\Sun}\, \omega_\text{Kepler}
	\eeq
	Here, $k_\Jupiter$ depends on the eigenvalues of the perturbing Hamiltonian, and $\omega_\text{Kepler}$ is the frequency of a Keplerian orbit.
		
	To evaluate this, we need expressions for Jupiter's location and mass\endnote{This frequency also depends nontrivially on the separation between Jupiter and Saturn, an effect which we ignore here.  For an interesting proposal to search for observational signatures of a selection effect on this front, see \citep{waltham2006large}}.  Our discussion assumes a system architecture similar to the solar system's, in which giant planets occur in the outer regions of the stellar system, with small planets in the interior.  Generically, the innermost giant planet is expected to be found at a few times the ice line, due to the buildup of icy material at that location \citep{ida2008toward}.  This is in line with exoplanet observations that find an increase in giant planet occurrence at this location, initially observed in \citep{cumming2008keck} with limited data, and more recently corroborated with more data in \citep{fernandes2019hints}.  The location of the ice line during planet formation depends on the dominant source of heating (irradiation or accretion), but it was found in our previous studies that the choice between the two sources did not affect multiverse probabilities significantly.  For the remainder of this section, we assume that irradiation is the dominant source of heating, which results in $a_\text{ice}=297.8 \lambda^{43/30}M_{pl}^{2/3}/(\alpha^4m_e^{4/3}m_p^{1/3})$ (see \cite{mc2} for details).
	
	The mass giant planets attain is somewhat less understood.  Many sources take the final mass to scale as the thermal mass or the pebble isolation mass \citep{bitsch2018pebble}, a stance we adopted in our previous papers.  However, a recent paper \citep{ginzburg2019end} argues that this limiting form is not appropriate, and instead that the growth of giant planets is governed by inviscid gap formation and disk lifetime, resulting in a final mass that scales as
	\beq
	M_\Jupiter\sim (\omega_\text{Kepler}t_\text{disk})^{10/147}
	\left(\frac{H_\text{disk}}{a}\right)^{134/49}
	\left(\frac{G\rho_\text{disk}}{\omega_\text{Kepler}^2}\right)^{1/3}M_\star
	\eeq
	Using the expressions for disk lifetime $t_\text{disk}$, height $H_\text{disk}$, and density $\rho_\text{disk}\sim\Sigma/H_\text{disk}$ found in the appendix, we find that Jupiter's mass depends on fundamental constants as
	\beq
	M_\Jupiter = 2.4\times10^4\,\frac{\kappa^{1/3}\,\lambda^{2.01}\,M_{pl}^{2.43}}{\alpha^{4.06}\,m_e^{.62}\,m_p^{.82}}
	\eeq
	and that the dominant secular frequency scales as
	\beq
	\Omega_\Jupiter=5.1\times10^{-3}\,\frac{\kappa^{1/3}\,\alpha^{2.47}\,m_e^{1.74}\,m_p^{.24}}{\lambda^{.81}\,M_{pl}^{.98}}
	\eeq
	In universes where $\Omega_\Earth>\Omega_\Jupiter$, a large moon is not required to stabilize the obliquities of moonless inner planets.  We now calculate the scale of $\Omega_\Moon$ in terms of fundamental constants.
	
	\subsection{What Sets the Size of and Distance to the Moon?}
	
	To determine the dependence of $\Omega_\Moon$ on physical constants, we need to understand the typical mass and distance scale involved in moon forming impacts.  Ideally, we would employ a distribution of moon properties derived from first principles to estimate the fraction of moons with a given precession frequency, but this distribution depends on the precise details of planet formation scenarios in addition to moon formation during impacts, and has yet to be fully elucidated \citep{carter2020energy}.  In the following, we find typical values for these quantities based off the scales involved in the setup, and argue that the resultant frequency can be treated as somewhat typical among planet-moon systems.
	
	It is by now well agreed upon that Earth's moon must have formed in a collision with a Mars sized planet relatively late in Earth's formation history, as originally put forth in~\citep{cameron1976origin}.  These giant collisions are expected to be generic, are thought to have affected many of the planets within the solar system (for a recent literature review see \citep{carter2020energy}), and have already been indirectly inferred in other star systems \citep{meng2014large}.  However, most collisions do not result in the formation of a bound moon; in \citep{stewart2012collisions} it was found that a relatively narrow window in collisional phase space is required to avoid either complete merger between bodies or a hit and run scenario.  Roughly, they find that impactor speed must be less than $1.1v_\text{esc}$, and that the impact parameter must be within $0.65-0.8 R_\Earth$.  The viability of an impact possessing these conditions will be dealt with below; here, we find the size and orbit of a moon that results from a collision that satisfies these properties.
	
	The typical separation that results from a giant impact that results in a bound moon was determined in \citep{canup2004simulations} in terms of the planet's Roche limit as $a_\Moon\sim 2.9 R_\text{Roche}$.  Since, as previously mentioned, the planet is assumed to be rotating at nearly its breakup speed, the Roche limit will be on the order of the planet size, and so $a_\Moon\sim\mathcal{O}(\text{few})\times R_\Earth$.
	
	The mass of the resulting moon can be determined by angular momentum conservation, but is essentially determined by the amount of the impactor that geometrically interacts with the planet, which will be $\mathcal{O}(1)$ \citep{leinhardt2011collisions}.  Then, the moon's mass is determined by the mass of the final giant impactor, which is given in terms of the isolation mass.  From the appendix, this depends on constants as $M_\text{iso}=6.3\times10^6 \kappa^{3/2} \lambda^{1/2} m_p M_{pl}^{3/2}a^{3/2}$.  The obliquity precession frequency induced by a planet's moon is
	\beq
	\Omega_\Moon = 2.4\times10^{-3}\,\frac{\kappa^{3/2}\,\lambda^{25/8}\,m_p^5}{\alpha^{15/2}\,m_e^{9/4}\,M_{pl}^{7/4}}
	\eeq

	\subsection{Are Gentle Collisions Generic?}
	
	We had noted above that moon forming collisions are only possible if the impact is `gentle' in the sense that $v_i\sim v_\text{esc}$.  This is not necessarily automatic, especially since Earth's escape velocity (11 km/s) is less than its Keplerian velocity (30 km/s).  The only reason a gentle collision was possible for the solar system's configuration is because most objects inherit their orbital characteristics from the primordial disk, and so typical impact speeds are reduced by a factor of eccentricity, so that $\bar v_i\sim\bar e\, v_\text{Kepler}$ \citep{matsumoto2017formation}.  If the typical impactor speed were much larger, then essentially all impacts would either be complete mergers or hit and runs, and large moons would consequently be exceedingly rare.
	
	Using Equation~(\ref{bare}) for the average planet eccentricity and specializing to temperate, terrestrial planets, we find the ratio of these two speeds can be expressed as
	\beq
	\frac{v_\text{esc}}{\bar v_i}=0.011\,\frac{\alpha^2\,\beta}{\kappa^{1/2}\,\lambda^{11/16}\,\gamma^{1/8}}
	\eeq
	In our calculations, we take gentle collisions to be possible only if this quantity is greater than 1, so that $f_\text{gentle}=\theta(v_\text{esc}-\bar v_i)$, where $\theta(x)$ is the Heaviside step function.  This discounts the possibility for some happenstance collision to be gentle enough in universes where this quantity is less than 1, rare as these may be.  Accounting for these rare cases in more detail does not affect our results.  
	
	In fact, this factor has almost no effect on the numerical values of the resulting probabilities.  So, even though the relation $v_i\sim v_\text{esc}$ may seem like an intriguing coincidence, the origin of this fact cannot be explained by any multiverse selection effect.  Neither does it seem that it can be explained by selection effects within our universe; the ratio of speeds scales as $v_\text{esc}/v_i\,\propto\, M_\text{planet}^{1/3}/(M_\star\,a)^{1/4}$.  Because this dependence is so weak, a planet would have to be 800 times smaller for the effect on moon forming collisions to become appreciable.
	
	\subsection{What Sets the Moon Retention Timescale?}
	
	An additional factor to take into consideration is whether planets are able to retain their moons for an appreciable amount of time.  In general, the moon retention timescale depends greatly on system properties such as moon size, initial separation, and dissipation factors \citep{tokadjian2020impact}.  However, using our estimates from above, we can derive the magnitude of the scales involved, and evaluate the typical timescale for a representative system.
	
	To study the orbital evolution of the moon, we use the constant phase lag model found in \citep{efroimsky2013tidal}, where the gravitationally induced bulge in the planet's ocean exists at a fixed phase from the moon's orbital location, inducing a tug that either leads to outspiraling or inspiraling, depending on whether the moon's orbital period exceeds or subceeds the planet's rotation, respectively.  This model can be integrated to yield the time it takes for the moon to attain any orbit as
	\beq
	t(a)=\frac{2}{39}\frac{Q}{k_2}\,\frac{M_\Earth^{1/2}\,R_\Earth^{3/2}}{G^{1/2}\,M_\Moon}\,\frac{a^{13/2}-a_0^{13/2}}{R_\Earth^{13/2}}
	\eeq
	Here $Q/k_2\sim40$ parameterizes the Earth's deformational response to tidal forces.  Here, we have assumed that the moon is drifting outward, but the opposite case may be accommodated by flipping the sign of this expression.
	
	The dimensional timescale in this expression is very short, close to a year.  The reason the moon has remained stable over the course of Earth's evolution stems from the steep dependence on distance, which causes a drastic slowdown in migration with orbit.  The full lifetime of the Earth-moon system (neglecting other factors like the finite lifetime of the sun) can be obtained as the point in time for which the moon exceeds a stable orbit, which is given in terms of the Hill radius as $a_\text{max}=0.5R_\text{Hill}$ \citep{domingos2006stable}.  With this, the orbital lifetime of a large moon around a terrestrial planet can be given in terms of fundamental constants as
	\beq
	t_\Moon = 2.0\times10^{-6}\,\frac{\lambda^{73/12}\,m_p^{19/4}}{\kappa^{3/2}\,\alpha^{37/2}\,m_e^{17/4}\,M_{pl}^{3/2}}
	\eeq
	This has been normalized to the pessimistic estimate of 68 Gyr for our values \citep{tokadjian2020impact}.  We include this effect in our habitability hypotheses as a factor $f_\text{ret}=\theta(t_\star-t_\Moon)$.
	
	\subsection{Does a Giant Impact Phase Always Occur?}
	
	A final consideration is whether a giant impact phase is a necessary outcome of planet formation.  Not only is this necessary for the formation of a large moon, but giant impacts were probably also necessary for early Earth to become hot enough to become molten; this in turn is a prerequisite for core differentiation \citep{kaula1979thermal}.  At the outset, it is unclear whether the giant impact phase is generic, or whether the few dozen Mars sized planets that originally result from planet formation could in principle be stable for different values of the physical constants.
	
	Determining whether a planetary configuration is stable is a difficult task.  Though at present only a full simulation can assess a given configuration with (near) perfect accuracy, several analytical methods have been developed.  The most ubiquitous is the angular momentum deficit criterion \citep{laskar2017amd}---this treats all perturbations as secular, and notes that a certain quantity, the angular momentum deficit, is conserved in this limit, and strictly decreases during mergers or ejections, resulting in greater stability.  If this quantity initially does not exceed some threshold, then the system will remain stable.  For the restricted three body system this results in a criterion for the eccentricities and orbital spacings of the two planets.  If these are taken to be equal mass, this gives a condition for the critical eccentricity:
	\beq
	e_\text{AMD}=\frac{\sqrt{1 - \frac{\Delta a}{a} + \left(\frac{\Delta a}{a}\right)^2}-1}{\frac{\Delta a}{a}-1}\approx\frac12\frac{\Delta a}{a}
	\eeq
	This approximation is always valid to within an error of 0.02.  From here, however, it is clear that the critical eccentricity scales in the same manner as the average eccentricity, since $\Delta a\sim R_\text{Hill}$.  Therefore, since the eccentricity distribution is set entirely by this value, the fraction of systems that exceed the critical eccentricity will be independent of physical constants.  If we take the results from \citep{quintana2016frequency} that $96\%$ of Earthlike planets experienced a giant impact, then we conclude that giant impacts are a generic feature throughout the multiverse.  We may extend this analysis to include other stability criteria, such as mean motion resonance overlap \citep{wisdom1980resonance}, or extensions thereof \citep{mustill2012dependence}, which are valid for more compact, strongly interacting systems.  However, these additional channels will always serve to make the systems more unstable, and so will not alter our conclusions.
	
	\subsection{Is a Large Moon Necessary for Complex Life?}
	
	We can finally put the pieces we have discussed together in a single expression for the fraction of planets with small, stable obliquities:
	\bea
	f_\text{ob}= \theta\left(\Omega_\Earth +\Omega_\Moon>\Omega_\Jupiter\right)\Big[f_\Moon\,\theta\left(\Omega_\Earth <\Omega_\Jupiter\right)
	+\theta\left(\Omega_\Earth >\Omega_\Jupiter\right)\Big]\label{fob}
	\eea	
	Here, $\theta(x)$ is the Heaviside step function, and we have defined $f_\Moon=0.014f_\text{gentle}f_\text{ret}$.  The coefficient is taken from \citep{waltham2011testing} as the fraction of giant impacts that result in a sufficiently large moon, though the exact number depends on the details of the planet formation scenario and is not known.  In principle, this expression for $f_\text{ob}$ could be integrated over the distributions of the relevant frequencies that result from a distribution of orbital elements and planetary characteristics; here, we treat these as only depending on stellar mass.
	
	With the habitability assumption that a large moon is necessary for complex life, the probabilities of observing our values of the constants are
	\bea
	\mathbb{P}(\alpha_{obs})=0.21,\,\, 
	\mathbb{P}(\beta_{obs})=0.02,\,\, 
	\mathbb{P}(\gamma_{obs})=0.05,\,\,
	\mathbb{P}(\delta_{u\phantom{!}obs})=0.27,\,\,
	\mathbb{P}(\delta_{d\phantom{!}obs})=0.37,\,\,
	\eea
	These are all quite reasonable, and not very different from the baseline probabilities displayed in Equation~(\ref{basep}). despite the numerous boundaries associated with obliquity stabilization.  However, the probability of being situated around a star at least as massive as the sun becomes $	\mathbb{P}(M_\Sun)=0.017$, as opposed to $0.24$ without including this criteria.  This is because obliquities of bare planets are stable around low mass stars, and so high mass stars like our sun are penalized by requiring an unlikely moon arrangement for habitability, consistent with the findings of \citep{deitrick2018exo}.  Note that this probability scales approximately linearly with $f_\Moon$, so this conclusion is contingent on our choice of estimate for fraction of planets with large~moon.
	
	Furthermore, the fraction of observers experiencing small $\Omega_\Earth$ but adequate $\Omega_\Moon$ are quite small, as indicated in Figure~\ref{omom}.  The confidence we assign to this depends on the precise statistical measure we use, but, for instance, we find that $\mathbb{P}(\Omega_\Earth<\Omega_\Jupiter\,\&\,\Omega_\Moon>\Omega_\Jupiter|\text{obs})=0.036$.  This can be contrasted with the case when restricting to observers in our universe, in which case this quantity equals 0.48.  From this, we conclude that the notion that large moons are essential for planetary habitability, through their effect on obliquity stabilization, is disfavored with the multiverse hypothesis.
	
	Again, we stress that we have restricted our analysis to the case where large moons are important because of their effect on obliquity stabilization.  If large moons are necessary for some other reason, this may well be compatible with the multiverse, and indeed taking $f_\text{ob}=f_\Moon$ rather than Equation~(\ref{fob}) yields results consistent with the multiverse.  However, additional factors must be included to properly evaluate the other proposed reasons why large moons are necessary.
	
		\begin{figure}[H]
			\includegraphics[width=0.85\textwidth]{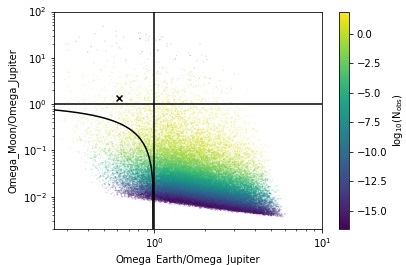}
			\caption{Ratio of frequencies for a sample of $10^5$ points.  The color denotes the number of observers for a given parameter choice (relative to our values).  The dashed line denotes the region of stable obliquities.  The cross marks the values of the Earth-Moon system.}
			\label{omom}
		\end{figure}
	
	\section{Water Delivery}\label{Water}
	
	An especially alluring habitability criterion is the presence of surface liquid water.  This notion is so ingrained in us that the conventional definition of a star's habitable zone is the region capable of hosting liquid surface water, and the unofficial astrobiology tagline is to `follow the water'.  Looking at different bodies within our solar system, and more recently, with the characterization of exoplanet systems, it has become clear that a planet's water content can be highly variable \citep{sing2016continuum}.  As such, making assumptions about the necessary water required for complex life has drastic implications for its expected distribution throughout the universe.  Furthermore, since the source of Earth's water is still contested, making assumptions about the delivery mechanisms contribute equally.  We evaluate both potential delivery scenarios as well as habitability criteria to determine which are compatible with the multiverse.
	
	The source of Earth's water is still unknown, as the cacophony of ongoing publications will attest.  The mystery arises because the Earth was presumably formed inside the water ice line (though see \citep{morbidelli2016fossilized}), meaning that the rocks it most likely formed from were drier than the Earth is observed to be \citep{morbidelli2012building}.  Much effort has been devoted to elucidating ways water could have later been delivered to the inner solar system, as well as ways water could have been present in the requisite amounts despite the high temperatures.  The water sources we consider are as follows: (i) delivery from planetesimals during giant planet formation, (ii) delivery from planetesimals during the grand tack, (iii) delivery from comets, and (iv)~oxidation of a primordial hydrogen atmosphere by a magma ocean.  Each of these has had varying degrees of success satisfying isotopic, elemental and solar system constraints, though we adopt an inclusive stance in the hopes of garnering further insight into which of these mechanisms are potentially viable.  It is important to note that no current theory explains all features of Earth's volatile composition completely satisfactorily \citep{meech2020origin}, giving reason to keep an open mind about the subject.
	
	This is not the setting to go into a great amount of detail about the various constraints on each of these scenarios, and so for further information the reader is referred to several excellent recent reviews on the topic: \citep{meech2020origin,dauphas2013geochemical}.  To frame our discussion, we briefly review some of the main constraints, and the implications they have for each scenario.  The primary constraint is the deuterium to hydrogen (D/H) ratio; Earth's is quite a bit higher than the sun's and the protoplanetary disk, and lower than the Jupiter family comet 67P \citep{altwegg201567p}.  Of additional importance is that if anything, Earth's D/H ratio increased with time due to atmospheric escape (by a factor of 2--9)\citep{genda2008origin}, and based off observations of lava, the deep mantle likely has a lower D/H \citep{hallis2015evidence}. 
	
	Noble gases provide good additional constraints on Earth's volatile accretion history because their inertness makes their dynamics relatively straightforward; generically, the Earth is depleted in noble gases \citep{zahnle2010earth} with respect to both solar and chondritic composition.  Earth's Ne and Ar ratios are consistent with a mixture of nebular gas and chondrites and minimal atmospheric escape, but a dearth in atmospheric Xe has remained challenging to explain \citep{marty2012origins}.  A recent study finds a consistent scenario where comets deliver 20$\%$ of Earth's noble gases, and negligible amounts of other volatiles, with EUV-ionized Xe preferentially escaping due to its relatively low binding energy \citep{bekaert2020origin}.
	
	Additionally, nitrogen provides insight into the origin of the Earth's atmosphere.  isotopic ratios provide independent evidence for meteoric delivery \citep{marty2012origins}.  One study recently reported that C/N ratios are consistent with a primordial nitrogen reservoir \citep{marty2020evaluation}, but another argues for delivery \citep{grewal2019delivery}.  Recent modeling of nitrogen sources and sinks on the early Earth suggest nitrogen partial pressure was considerably lower than that of today, suggesting later delivery \citep{gebauer2020atmospheric}.  Generically, Earth's C/N is high compared to chondrites, though a recent explanation in terms of a giant impact was given in \citep{grewal2019delivery}.
	
	Finally, transition metals provide additional clues to the source of volatiles.  O, Cr, Ti, and Mo are most consistent with an enstatite chondrite composition for Earth \citep{drake2005origin}, but these are highly reduced, and thus too dry.  Though there is a possibility that these were initially wetter and have since desiccated, this suggestion does not explain why some asteroids have remained wet \citep{drake2005origin}.
	
	Each scenario will dictate the amount of water delivered to Earth in terms of physical constants.  The habitability conditions we consider will all be functions of this amount.  We estimate this quantity for each scenario first, and then synthesize them into our habitability conditions afterward.
	
	\subsection{Asteroid Injection}
	
	The first scenario we consider, and also currently the most popular, is water delivery via asteroids early in the solar system's history.  There are two versions of this scenario: delivery during the formation of the giant planets, and during an instability such as the grand tack scenario.  The first is more generic, as grand tacks may be rare, while the second is capable of delivering more material.  Uncertainties in the resulting total mass injected make it currently unclear whether the former scenario is capable of delivering sufficient water to explain Earth's water content, or whether the latter scenario is required.  So, we consider each in turn.
	
	As explored in detail in \citep{raymond2017origin}, giant planet formation is unavoidably accompanied by the injection of planetesimals into the inner system, because rapid growth creates a gravitational perturbation that destabilizes nearby bodies.  In this scenario, the total mass that gets delivered to a terrestrial inner planet is given by the total amount of mass scattered inward $M_\text{scatter}$, times the fraction of that mass which is ice $f_\text{ice}$, times the fraction of bodies that do not get orbitally damped by the time they enter the inner system $f_\text{undamped}$, times the fraction that hit Earth $f_\text{hit}$, giving
	\beq
	\mwater=M_\text{scatter}\,f_\text{ice}\,f_\text{undamped}\,f_\text{hit}
	\eeq
	We will determine the scaling of each of these factors with physical constants, and so how this quantity changes in other universes.
	
	The scattered mass is given by $M_\text{scatter}\sim \pi\Sigma_p(a)a\Delta a$, where $\Sigma_p$ is the planetesimal surface density, and $\Delta a$ is the region within the disk that becomes unstable as a result of growth.  The surface density was found in \citep{mc2} to be given by $\Sigma_p\sim 0.01ZM_\star /(r_\text{disk}a)$, where $Z\sim0.02$ is the metallicity and $r_\text{disk}$ is the radius of the disk.  This is to be evaluated at Jupiter's orbit, which as explained above is given as a few times the ice line.  The width of the instability region is of the order of its location; from \citep{raymond2017origin}, a hard inner edge is $a_\Jupiter/2$, where every eccentricity less than 1 is stable, running through several times $a_\Jupiter$ as seen from simulations.  The formation of the other giant planets give subdominant contributions, but scale the same as above.  Though the factor $f_\text{ice}$ will depend on orbit and is key in differentiating asteroids from comets, we hold it fixed here, as generically we expect asteroids to originate around the ice line \citep{martin2012formation}.
	
	If a planetesimal does get destabilized, gas drag may damp its orbit before it reaches the inner solar system, emplacing it in the asteroid belt (or, if even faster, circularizing the orbit before appreciable scatter).  This effect is size dependent, so that more massive objects are more likely to remain undamped until they reach the inner system.  An estimate for the damping rate is obtained from the drag equation $F\sim \rho_\text{gas}A v^2$.  This gives a damping time
	\beq
	t_\text{drag}\sim \frac{\rho_\text{rock}\,H_\text{disk}\,d_\text{planetesimal}}{\Sigma\,v}
	\eeq
	For solar system values, this yields $t_\text{drag}\sim 10 (d/\text{km})(a/\text{AU})^{21/8}$ yr.  This can be used to determine the smallest body that will remain undamped by comparing the drag timescale to the orbital timescale.  Combining this with the planetesimal size distribution yields the fraction of bodies that remain undamped until the inner system.
	
	The planetesimal size distribution is not agreed upon, but is generically expected to resemble a power law, $dN/dM\propto M^{-p}$.  A standard value is \emph{p} = 1.6 from \citep{simon2016mass}, though it was determined in \citep{mc2} that precise choice of p does not affect results much.  With this distribution, then the fraction of mass delivered before damping is given by
	\beq
	f_\text{undamped} =R\left(1-\left(\frac{M_\text{damp}}{M_\text{max}}\right)^{2-p}\right)
	\eeq
	Here, $R(x)=0$ for $x<0$ and $R(x)=x$ for $0<x<1$, and this formula is valid for any $p<2$.  This requires a maximal planetesimal mass, for otherwise the distribution used yields infinite total mass; for this, we use the isolation mass at the specified orbit.  To evaluate this, we need to specify the ratio $M_\text{damp}/M_\text{max}$.  Unless otherwise specified, we take this to be 0.5 for our constants and around a solar mass star, corresponding to an $f_\text{undamped}=0.24$.  In Figure~\ref{fdamp} we display how the probabilities depend on this parameter.
		\begin{figure}[H]
			\includegraphics[width=0.96\textwidth]{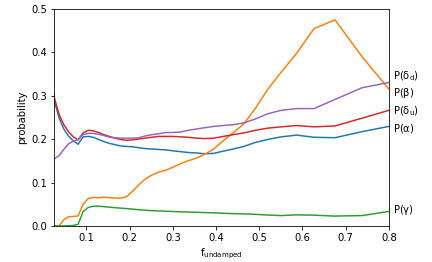}
			\caption{Dependence of the probabilities of observing our constants as a function of the fraction of disrupted planetesimals that enter the inner solar system, as opposed to becoming emplaced in the asteroid belt.}
			\label{fdamp}
		\end{figure}
	
	The last quantity, $f_\text{hit}$, is on the order of $0.4$ \citep{o2006terrestrial}.  An order 1 fraction of planetesimals scattered in to 1 AU will end up hitting the Earth because they are progressively less likely to get kicked further in, and are likely corotating close to the ecliptic.
	
	All told, this gives the average water fraction of terrestrial planets as
	\beq
	\fwater = 12.7\,\frac{\kappa\,\lambda^{21/10}\,\gamma^{1/3}}{\alpha^{11/2}\,\beta^{25/12}}\,f_\text{undamped}
	\eeq
	Throughout, we normalize this quantity to $\fwater=2.5\times10^{-4}$ for our values to match Earth's ocean mass.  We will discuss mantle water below, but note here that this quantity is ultimately translated to surface area fraction, which is done in a manner that remains agnostic to the total amount of subsurface water.  It should be noted that the distribution of water fractions in this scenario is broad because the delivery is dominated by the largest bodies, making the outcome stochastic \citep{bottke2010stochastic}.
	
	\subsection{Grand Tack}
	The viability of the previous scenario is dependent on the total mass of planetesimals delivered to the inner system, which is imprecisely known.  Outer systems that undergo an initial instability, as has been proposed to have happened in our solar system, have the potential to deliver more water, but are correspondingly rarer.  Here we explore the consequences of adopting the hypothesis that grand tack instabilities are required to deliver the requisite amount of water to inner planets, and estimate the fraction of systems that undergo such an evolution.
	
	\textls[-5]{The grand tack was initially proposed to explain the smallness of Mars and the emptiness of the asteroid belt \citep{walsh2011low}.  In this scenario, Jupiter forms rapidly while still embedded in the protoplanetary disk, and undergoes migration inward, clearing out material between its formation location and 1 AU.  After somewhat of a delay, Saturn undergoes runaway growth as well, halting Jupiter's inward migration and then pulling it back outward.  In addition to clearing out the original asteroid belt, this migration scatters wet material inwards, resulting in enough water delivered to account for Earth's budget~\citep{o2014water}\endnote{A resonant instability that also is purported to have occurred in our system, a la the Nice model, is less suited to explaining the source of Earth's water, as it happened late enough to disagree with isotopic \citep{greenwood2018oxygen} as well as zircon \citep{wilde2001evidence,mojzsis2001oxygen} constraints.}.}
	
	The dynamics of this process are similar to the growth destabilization scenario considered above, and so the amount of material scattered inward can be treated as nearly identical to before (though it will deliver more, the scaling laws are identical).  Thus, the only difference arising from adopting this hypothesis comes from the factor $f_\text{tack}$, the fraction of planetary systems undergoing a grand tack scenario.  As may be expected from such a baroque evolution, grand tacks are the exception in planet formation scenarios, which could also have resulted in a full migration inward to form hot Jupiters \citep{goldreich1980disk}, or else almost no orbital migration, as in the classical formation scenario.  The fact that Jupiter migrated from about 3.5 AU to 1.5 AU before Saturn formed and pulled it back out is due to the fact that $t_\text{migration}\sim t_\text{formation}$.  However, since these quantities depend on constants in different ways, we can expect that this rough equivalence does not hold throughout the multiverse.
	
	The type II migration timescale for gas giants was found in \citep{mc2} to depend on physical constants according to  $t_\text{migration}\sim10^7\lambda^{73/60}m_p^{1/3}M_{pl}^{5/6}/(\alpha^4m_e^{13/6})$.  The gas giant accretion timescale can be given in terms of the Bondi growth time, $t_\text{Bondi}=M/\dot{M}=c_s^3/(4\pi G^2\rho M)$~\citep{ginzburg2019end}.  Here, runaway growth occurs on a timescale dictated by the initial mass, which we take to be the isolation (thermal) mass at several times the snow line.  Then we have
	\beq
	t_\text{Bondi}=1.1\times10^{-9}\,\frac{m_e^{5/3}\,M_{pl}^{2/3}}{\kappa^{5/2}\,\lambda^{7/30}\,m_p^{10/3}}
	\eeq
	This scales with semimajor axis as $t_\text{Bondi}\propto a^{1/2}$.
	
	The formation of gas giants will follow some distribution around this typical value.  Here, we can take formation times to be exponentially distributed.  In this case, the time lag between the formation of Jupiter and Saturn $\Delta t$ will have a cumulative distribution 
	\beq
	c_\text{delay}(\Delta t)=1-k_\text{delay}\,e^{-\Delta t/t_\Jupiter}
	\eeq
	valid for $\Delta t>0$, which will be enough for our purposes, though there may be some systems where the outermost planet forms first.  The prefactor $k_\text{delay}$ will be $1/2$ if we neglect any difference in formation times between the two planets, and will be $0.41$ if we take $a_\Saturn=2a_\Jupiter$.  This overall coefficient will have no effect on our probability calculations, as it just amounts to a constant rescaling, and therefore has no bearing on relative probabilities.  Additionally, we have checked that taking planet formation times to be uniform rather than exponentially distributed does not significantly affect our probability estimates.
	
	This can be turned into a distribution of migration distances, if the migratory evolution is specified.  For this, we use the formalism in \citep{crida2007cavity}; they have that the torque on the planet $\tau\,\propto\, a^{3/2}\Sigma(a)\nu(a)$, and angular momentum $L(a)\,\propto\, a$, giving upon integration of $\dot L=\tau$
	\beq
	a(\Delta t)=\frac{a_0}{\left(1+\frac{\Delta t}{t_{II}}\right)^{4/3}}
	\eeq
	As in \citep{walsh2011low}, we take $a_0=3.5$ AU.  In their scenario, Jupiter migrates until it reaches 1.5~AU, corresponding to $\Delta t=0.89\,t_{II}$.  To determine the fraction of planets that undergo a grand tack, we take the outer migration boundary to be 2.7 AU, when Jupiter crosses the asteroid belt, corresponding to $\Delta t=0.21\,t_{II}$.  The inner boundary is given by the habitable zone 1 AU, beyond which any initial temperate planets would be scattered or worse.  This corresponds to $\Delta t = 1.56\, t_{II}$.  The fraction of systems that experience a grand tack is
	\bea
	f_\text{tack} = k_\text{delay}\left(e^{-0.24z}-e^{-1.76z}\right),\quad
	z=1.3\times 10^{16}\,\frac{\kappa^{5/2}\,\lambda^{29/20}}{\alpha^4\,\beta^{23/6}\,\gamma^{1/6}}
	\eea
	
	Incidentally, we consider the habitability hypothesis that depends only on a grand tack being present, irrespective of whether it is the main source of water delivery.  This may be motivated, for example, by the expectation that grand tacks could be responsible for clearing out the asteroid belt, and this could conceivably be the origin of the coincidence between timescales.  For this, the {\bf yellow + entropy + temp + terr + C/O + grand tack} condition, we find
	\bea
	\mathbb{P}(\alpha_{obs})=0.20,\,\, 
	\mathbb{P}(\beta_{obs})=0.02,\,\, 
	\mathbb{P}(\gamma_{obs})=0.02,\,\,
	\mathbb{P}(\delta_{u\phantom{!}obs})=0.27,\,\,
	\mathbb{P}(\delta_{d\phantom{!}obs})=0.40,\,\,
	\eea
	These are slightly worse, though not wholly different from, the baseline probabilities in Equation~(\ref{basep}), and so we conclude that the presence of grand tacks is not a major factor in explaining our presence in this universe.
	
	\subsection{Comets}
	
	Comets are the originally proposed source of Earth's water \citep{delsemme1992cometary}.  They are distinct from asteroids as being further out in the solar system, where temperatures below freezing points lead to a much higher volatile content, and as such suggest themselves quite naturally.  However, this explanation of Earth's water has been challenged in recent years from isotopic considerations, with limits of varying degree being placed on the total allowable amount of water delivered from this source: \citep{drake2002determining} claim that <$50\%$ of Earth's water can be cometary in origin, based off D/H ratios.  \citep{dauphas2000late} has that number as <$10\%$ based off the same consideration, and \citep{bar1998trapping} finds <$1\%$ of material delivered can be cometary based off noble~gases.
	
	Nevertheless, there are still several caveats in these limits that make a cometary origin of Earth's water hypothetically viable.  Firstly, the isotopic composition of comets is inferred from only a few samples, in which there is significant spread; it is possible that we have only measured values in outliers, and that the majority of comets (in the early system) more closely align with Earth's isotopic composition.  Secondly, some comets, such as Halley \citep{eberhardt1988d} do match Earth's D/H ratio.  Additionally, cometary noble gas abundance measurements have been very limited, with limits only stemming from inferred properties \citep{owen1992possible}.  Lastly, cometary abundances are mostly measured from their outgassed tails, which may not perfectly reflect the bulk composition \citep{drake2005origin}.  These caveats cast enough allowance on this mechanism to warrant an exploration of whether it is compatible with the multiverse.
	
	Like the case of asteroids above, the total amount of water delivered by this mechanism can be given by the total mass present in icy bodies initially, multiplied by the fraction that arrive on Earth:
	\beq
	\mwater = M_\text{comets}\,f_\text{ice}\,f_\text{hit}
	\eeq
	Here, the total cometary mass is given roughly by the amount of material in the outer system, $M_\text{comets}\approx\pi\Sigma(r_\text{disk})r_\text{disk}^2$.  This material gets scattered during the evolution of the outer system, the majority being lost to space or flung into the sun, a small fraction emplaced in the Oort cloud, and an even smaller fraction delivered to planets.  The fraction that is delivered to Earth can be found from the ratio of gravitational cross sections of the Earth and sun, $f_\text{hit}=\sigma_\Earth/\sigma_\Sun=R_\Earth^2/(2R_\Sun a_\text{temp})$, in agreement with simulations \citep{fi}.  This fraction is significantly smaller than the asteroidal case, where injected material is perturbed comparatively slightly, giving preference to wide perihelia on roughly corotating~orbits.
	
	The water fraction resulting from this delivery mechanism is then
	\beq
	\fwater = 3.9\times10^{-5}\,\frac{\alpha^{9/2}}{\lambda^{31/20}\,\beta^{1/4}\,\gamma^{1/2}}
	\eeq
	This has been normalized to match Earth's ocean mass.  Though the uncertainties in the amount delivered range by several orders of magnitude, there is no obstruction to delivering the requisite amount of water by this method.
	
	\subsection{Magma Ocean}
	
	One further plausible source of Earth's water is the oxidation of a primordial hydrogen atmosphere from an initial magma ocean, as proposed in \citep{sasaki1990origin}.  In this scenario, oxygen contained within the mantle spontaneously reacts with the atmosphere to form water, and a magma ocean is essential to provide a continuous supply, preventing the surface from becoming saturated.  Again, the main problems with this model stem from isotopic constraints; the initial D/H would be solar rather than chondritic.  While hydrodynamic atmospheric escape is able to increase the initial value to be more in line with observations~\citep{genda2008origin}, the timescale for this to occur is at odds with isotopic ratios of noble gases \citep{dauphas2013geochemical}.
	
	The resultant water content in this scenario is dictated by the size of the primordial atmosphere, as outlined in \citep{ikoma2006constraints}.  This follows from three assumptions: mantle oxygen does not act as the limiting factor, that the overturn timescale is very fast compared to the atmosphere lifetime, and that the protoplanetary disk lifetime is long compared to the atmospheric equilibration time.  All of these are satisfied by orders of magnitude within our universe, and hold over the entire relevant range of parameter space.
	
	The total water content that results is set by the equilibrium conditions, which relate the partial pressures of water vapor and hydrogen gas.  The equilibrium depends on the dominant oxide comprising Earth's mantle, so that the ratio of the two gases ranges between 0.88 and 24 \citep{robie1978thermodynamic} as long as some iron is present (but drops precipitously in its absence).  At our level of analysis, however, we simply treat the two abundances as roughly equal, $P_{H_2}\sim P_{H_2O}$. 
	
	It remains to estimate the total mass of the primordial atmosphere, to determine the final oceanic mass.  The isothermal atmospheric density profile of a planet embedded in a gaseous nebula is given by \citep{ikoma2006constraints}
	\beq
	\rho(R)=\rho_\text{disk}\, e^{\frac{R_\text{Bondi}}{R}-1}
	\eeq 
	Here, $R_\text{Bondi}=2/7 \,GM_\Earth/c_s^2$ is the Bondi radius, where the escape velocity matches the sound speed of the nebular gas, and represents the outer edge of the atmosphere, where gas density smoothly connects with the nebula.  For Earth, $R_\text{Bondi}\sim 16 R_\Earth$.  This discussion also assumes that the Bondi radius is less than the Hill radius, as otherwise that would effectively be the atmospheric outer edge.  This is also the case over the entire relevant range of parameter space.  This does not take into account additional hydrogen that is outgassed from the mantle, but any supplied from this source is likely to equilibrate quickly, leading to the same atmospheric profile \citep{katyal2020effect}.
	
	The total atmospheric mass can be found by integrating this density from the surface to the Bondi radius.  This gives
	\bea
	M_\text{atm}&=&4\pi\,\rho_\text{disk}\,R_\text{Bondi}^3\,F_\text{neb}\left(\frac{R_\Earth}{R_\text{Bondi}}\right)\nonumber\\
	F_\text{neb}(q)&=&\frac23-\frac{1}{6}e^{1/q-1}q\left(1+q+2q^2\right)
	+\frac{1}{6e}\left(\text{Ei}\left(1/q\right)-\text{Ei}(1)\right)
	\eea
	where $\text{Ei}(q)$ is the exponential integral.  For small values of $R_\Earth/R_\text{Bondi}$, as is appropriate for the Earth system, $F_\text{neb}(q)\rightarrow q^4 e^{1/q-1}$.  This illustrates an exponential dependence on planet mass, indicating a strong sensitivity for atmospheric mass.  This is in line with the observed sharp dichotomy between terrestrial planets with negligible atmosphere and gaseous planets with atmospheric mass rivaling rocky mass, with transition around few $M_\Earth$ \citep{fulton2017california}.  Concomitantly, we expect the ocean mass to depend sensitively on planetary, stellar and disk properties in this scenario.  This is akin to the strong variability of water mass found in the other scenarios as well, the difference here being that the final mass is deterministic, rather than stochastic.
	
	To get an estimate for dependence of ocean mass on fundamental constants, we restrict our discussion, as before, to temperate, terrestrial planets.  Then the Bondi mass can be estimated using $c_s^2\sim T_\text{irr}/m_p$, and the nebular density by $\rho_\text{disk}\sim \Sigma/H_\text{disk}$ to give
	\beq
	\frac{R_\Earth}{R_\text{Bondi}}=0.46\,\frac{\alpha^{3/4}\,\beta^{1/2}}{\lambda^{19/80}\,\gamma^{1/8}}
	\eeq
	and
	\beq
	\fwater = 5.2\times10^7\,\frac{\kappa\,\alpha^{47/8}}{\lambda^{1141/480}\,\beta^{1/2}\,\gamma^{13/16}}\,F_\text{neb}\left(\frac{R_\Earth}{R_\text{Bondi}}\right)
	\eeq
	
	With expressions for $\fwater$ for each of the four different water delivery scenarios, we can see how each affects the probabilities of observing our constants\endnote{In passing, we remark on one last scenario, that of water delivered by interstellar grains.  Precluding a scenario where the feeding zone of early Earth is greatly enhanced, as in \citep{raymond2006high}, this is usually not considered a viable option, because grains, and subsequently rocks, situated at 1 AU are extremely dry.  However, it was suggested in \citep{king2010computer} that these typical arguments are based off the assumption that grains are spherical, and that if instead their fractal geometry is taken into account, a much greater amount of water may adsorb onto the surface, enhancing the amount delivered.  Additionally, ref. \citep{nakauchi2021formation} find that proton irradiation on interstellar grains can form water by breaking silicate bonds. The final ratio depends on the grain size distribution, but more so on the adsorption rate, which scales as $e^{E_\text{mol}/T}$.  This factor is identical to the exponential dependence for the magma ocean scenario, where both the size and location of the Earth are dictated by $E_\text{mol}$.  As such, the probabilities in the grain adsorption scenario will be close, though not strictly identical to, the magma ocean scenario.}.  However, first we need to determine how habitability depends on water content.
	
	\subsection{How Does Habitability Depend on Water Content?}
	
	As established above, Earth's ocean is about $2.5\times 10^{-4}M_\Earth$.  This amount covers $71\%$ of the Earth's surface with ocean, the remainder being land.  That such an equitable partition arises from such an apparently arbitrary hierarchy between planet mass and ocean mass is quite a striking coincidence; it is easy to image that Earth's water content could have been an order of magnitude greater or smaller, resulting in either a completely wet or completely dry planet.
	
	There have been a number of previous studies addressing Earth's unusual water content.  In \citep{simpson2017bayesian}, it was argued that this arrangement is generically expected to be observed, if the distribution of water contents has a steep preference for larger values, and if waterworlds are uninhabitable.  This is in line with the analyses of \citep{foley2015role}, which finds that worlds with no surface land will not possess a climate-stabilizing weathering feedback loop capable of maintaining a temperate climate over geologic time.  Similarly, in \citep{lingam2019dependence} it was argued that habitability is greatest on planets with intermediate area fractions such as ours, as the total flux of eroded material is proportional to the amount of precipitation, which scales with ocean area, and the amount of exposed land.  In \citep{abbot2012indication}, a dynamical explanation is put forward; a waterworld may undergo a runaway greenhouse state, successively building up a water vapor atmosphere and losing it to space, until a significant amount of land is uncovered to undergo a stabilizing climate feedback.
	
	\textls[-15]{As the habitability criteria we consider are all functions of the fraction of a planet's surface area covered by ocean $f_w$, we must first relate this to the mass fraction \mbox{$\fwater=M_\text{ocean}/M_\Earth$}} we have been dealing with up to this point.  From this perspective, the reason the ocean only partly covers the surface is due to the fact that the average ocean depth is similar to the scale of topographic variability.  As described in \citep{PL,mc3}, the topographic height scale, roughly equal to the average mountain height, is set by the condition that the gravitational and molecular forces on a mountain are balanced, yielding $h_\text{mountain}=0.0056M_{pl}/(\alpha^{1/2}m_e^{3/4}m_p^{5/4})$.  For Earthlike (temperate and terrestrial) planets, mountain height is proportional to planet radius, which is determined by a similar balancing condition.  In this case, the ratio of the ocean volume to the mountainous terrain volume is $V_\text{ocean}/(h_\text{mountain}A_\Earth)\propto \fwater$.
	
	The above discussion treats all of a planet's water as present in the ocean, though a considerable amount will be sequestered in the planet's interior (see \citep{ohtani2020role} for a recent review).  Though mantle water content depends on planet composition, temperature, size, age, and tectonic regime, the water content is never found to differ significantly from a few times the ocean mass \citep{hirschmann2006water}.  This is because, if both outgassing and subduction are proportional to the amount present in each reservoir, and the overall rates are set by the same convective processes, then the system will naturally equilibrate to maintain reservoirs of comparable size.  For our purposes, then, we take the amount of ocean water to be a constant rescaling of the total delivered water, and treat the details of this partitioning to not depend greatly on physical constants.
	
	Relating this to the fraction of area covered by ocean is a nontrivial task, and requires a hypsometric curve, which is the distribution of altitudes on a planet.  While for many planets this is expected (and observed) to be approximately given by a Gaussian distribution, Earth's height distribution is bimodal, as a consequence of its dual mafic/felsic composition~\citep{coradini1980hypsometric}.  This in turn is a consequence of Earth's abundant water, which mixes with original dense rock when subducted to create lighter continental shelves \citep{campbell1983no}.  
	
	Generically, if the fraction of land below a given altitude is given by $f_\text{hyp}(h)$, then the total volume of ocean water on a planet with sea level $h_\text{sea}$ is
	\bea
	V_\text{ocean}&=&\int^{h_\text{sea}}_0dh\, A_\Earth\, f_\text{hyp}(h)\nonumber\\
	&\equiv& h_\text{mountain}\,A_\Earth\, g_\text{hyp}(h_\text{sea})
	\eea
	where we have defined $g_\text{hyp}(h)=\int^h_0dh/h_\text{mountain}f_\text{hyp}(h)$.  Then the fraction of the planet's surface covered by water with a specified ocean volume is given by
	\beq
	f_w=f_\text{hyp}\left(g_\text{hyp}^{-1}\left(\frac{V_\text{ocean}}{h_\text{mountain} A_\Earth}\right)\right)
	\eeq
	For our purposes, we will follow \citep{simpson2017bayesian} and approximate Earth's hypsometric curve as a Gaussian, which will be adequate to capture the salient features relating ocean mass to area.  Even for this distribution, this expression is analytically intractable, but can be approximated as $f_\text{hyp}(g_\text{hyp}^{-1}(x))\approx 1-e^{-\sqrt{\pi}x}$, which captures appropriate asymptotics and is valid to within 0.02 everywhere.  Normalizing to reproduce the value $f_w=0.71$ for Earth's ocean content, we arrive at
	\beq
	f_w=1-e^{-4951.5\fwater}
	\eeq
	
	We may now compare various expectations for how habitability depends on water surface fraction that have been expressed in the literature, along with the multitude of water origin scenarios.  In \citep{lingam2019dependence}, the expectation is that biological productivity is proportional to both ocean area and land area, the former controlling the rate of evaporation and thus precipitation, the latter controlling the rate of erosion.  So, in this scenario, we find the habitability $\mathbb H$ to be $\mathbb{H}\,\propto\, f_w(1-f_w)$.  Alternatively, they consider that the emergence of complex life should be proportional to some greater powers of each of these quantities, if given by a sequence of hard steps, as in \citep{carterbio}.  For this scenario, they find $\mathbb H \,\propto\, f_w^6(1-f_w)^3$.  
	
	In \citep{simpson2017bayesian}, on the other hand, preference is given to dryer planets by weighting the purported fecundity of a planet by the amount of habitable land area squared.  While they consider several models for how the fraction of habitable (non-desert) land depends on water abundance, we only consider their extreme scenario where the two are independent, as the others closely resemble the scenarios considered above.  Then, we find $\mathbb H\,\propto\, (1-f_w)^2$.  As a counterbalance, we also consider the opposite preference, $\mathbb H\,\propto\, f_w^2$.  Though this has not, to our knowledge, appeared explicitly in the literature, it represents a possible expectation that worlds with higher water content may be more habitable.  In fact, it was found in \citep{kite2018habitability} that waterworlds can potentially have longer temperate durations than their Earthlike counterparts, as ocean pressure can act to regulate seafloor weathering.  In contrast, it was found in \citep{abe2011habitable} that desert planets have a larger habitable zone, as a consequence of their decreased propensity to undergo runaway greenhouse and icehouse phases.
	
	Finally, we consider simple cutoffs, between which habitability is insensitive to water content, but beyond which life is impossible.  There is, of course, considerable uncertainty as to which values of these cutoffs should be chosen, but we settle for representative values to demonstrate the effects these have on each delivery scenario.  The first is an upper cutoff, taken here to be twice the water volume, which for our hypsometric curve corresponds to $92\%$ ocean surface coverage.  For this, the habitability is $\mathbb H\,\propto\,\theta(0.92-f_w)$.  For a lower bound, we consider the point where mean annual precipitation falls below the minimum required for the sustenance of vegetation, as estimated in \citep{noy1973desert} from observations of desert ecosystems on Earth, and yielding $\mathbb H\,\propto\,\theta(f_w-0.05)$.  This neglects the inevitable evolutionary pressures life on purely desert worlds would face to allow greater biological productivity in arid climates, so this lower cutoff should serve as an illustrative stand-in.
	
	We take each of these six habitability hypotheses in turn, along with the four water delivery scenarios.  As a metric to encapsulate the validity of each combination, we take the naive Bayes factor relative to the null hypothesis, where habitability is independent of water fraction.  This is calculated as the product of observing the five physical constants we consider.  The results for this are reported in Table \ref{watertable} for the {\bf yellow + entropy + terr + temp + C/O} condition and  Table \ref{watertable2} for the {\bf nitrogen + area + Mg/Si} condition, the two conditions that give the highest probabilities found in \citep{mc5}.
	
	\begin{table}[H]
		\caption{Bayes factors for various combinations of water delivery mechanisms and habitability conditions.  All values are the product of the probabilities of observing the five values of the constants we vary, relative to the baseline case where habitability is independent of water content.  Values above $0.1$ are displayed in bold.  All values are computed assuming the {\bf yellow + entropy + terr + temp + C/O} condition.}
		\label{watertable}
		\setlength{\tabcolsep}{5mm}
		\begin{tabular}{ccccc}
				\toprule
				\boldmath{$\mathbb H$} & \textbf{Asteroids} & \textbf{Grand Tack} & \textbf{Comets} & \textbf{Magma Ocean} \\
				\midrule
				$f_w(1-f_w)$ & \bf 0.36 & \bf 4.6 & \bf 4.2 & \bf 1.4\\
                $f_w^6(1-f_w)^3$ & \bf 0.52 & \bf 11.0 & \bf 3.3 & \bf 1.4\\
                $(1-f_w)^2$ & \bf 0.44 & 0.067 & \bf 1.2 & \bf 0.12\\
                $f_w^2$ & \bf 0.51 & \bf 5.1 & \bf 0.56 & \bf 0.26\\
                $\theta(0.92-f_w)$ & \bf 0.94 & \bf 0.18 & \bf 2.5 & \bf 0.7\\
                $\theta(f_w-0.05)$ & \bf 0.56 & \bf 5.1 & \bf 1.0 & \bf 0.97\\
				\bottomrule
			\end{tabular}

	\end{table}
	\unskip
	
	\begin{table}[H]
\setlength{\tabcolsep}{5mm}
		\caption{Same as above, but with the {\bf nitrogen + area + Mg/Si} condition. Values above $0.1$ are displayed in bold.}
		\label{watertable2}
		\begin{tabular}{ccccc}
				\toprule
						\boldmath{$\mathbb H$} & \textbf{Asteroids} & \textbf{Grand Tack} & \textbf{Comets} & \textbf{Magma Ocean} \\
				\midrule
				$f_w(1-f_w)$ & 0.044 & \bf 0.46 & \bf 1.9 & \bf 0.14\\
                $f_w^6(1-f_w)^3$ & 0.055 & \bf 0.42 & \bf 1.6 & \bf 0.63\\
                $(1-f_w)^2$ & \bf 0.82 & \bf 0.58 & \bf 0.41 & 0.0013\\
                $f_w^2$ & \bf 0.24 & \bf 1.6 & \bf 0.77 & \bf 0.51\\
                $\theta(0.92-f_w)$ & \bf 0.93 & \bf 0.65 & \bf 2.1 & \bf 0.1\\
                $\theta(f_w-0.05)$ & 0.068 & \bf 0.78 & \bf 1.0 & \bf 0.99\\
				\bottomrule
			\end{tabular}
	\end{table}

	There are several important features that should be noticed about these tables.  The first is that certain combinations of habitability hypotheses and delivery mechanisms are highly disfavored in the multiverse scenario.  The second is that the specific combinations are sensitive to additional habitability assumptions that are made.  With the {\bf nitrogen + area + Mg/Si} condition, the intermediate water hypotheses of the first two rows are disfavored with an asteroid origin of Earth's water, and the grand tack and cometary scenarios generically do best across the board.  The magma ocean scenario fares best with habitability hypotheses that favor wetter planets.  Additionally, we should note that some of these habitability hypotheses are taken to be proportional to powers of $f_w$ and $(1-f_w)$, whereas in the original motivations these were accompanied by powers of $A_\text{planet}$.  In this regard, particularly the second hypothesis resembles a hard step process, which is extremely disfavored by our multiverse calculations; because they exhibit such a strong preference for large area, it is not uncommon for some of the probability values to be $\sim$$10^{-15}$, similar to the results found in \citep{mc3}.
	
	\section{Discussion}\label{Disc}
	
	We have considered several aspects of planetary habitability, and how these fare in terms of the multiverse framework.  Our discussion has necessarily been incomplete, as there are many facets to this problem that must be saved for future work.  We have focused on two: orbital habitability and water content.  Both of these present different general lessons for multiverse reasoning.
	
	Planetary habitability is rife with subtleties, as even confined to our own universe, selection effects play a large role in explaining why we are in this particular location.  Disentangling the selection effects inherent in our universe from the variation in planetary characteristics throughout the multiverse is challenging, and requires a thorough investigation of many subtle effects for each habitability condition.  However, once this is done, we are capable of shedding extra light on which habitability criteria are reasonable, and which are not.  Thus, it was only by determining that obliquities are stable throughout most other universes that we predicted that stable obliquities are not required for habitability.
	
	The mechanism of water delivery provides an interesting variation to the usual logic we employ.  For most of the habitability conditions we consider, the recommendation we make to test the theory follows the template of determining locations throughout our universe that in some sense mimic other universes, and checking whether those locales can or cannot support life.  For the water delivery scenario, the way we can test multiverse predictions is to understand more thoroughly the pathways via which water was delivered to Earth, and does not require a life hunting expedition.  As we account for possible habitability factors more thoroughly, we are beginning to uncover a clearer picture of which aspects of our universe are unique, which perceived oddities are meaningful, and which purported prerequisites for life are sensible.\\
	
\authorcontributions{Conceptualization, all authors; 
Methodology, M.S.; 
Formal Analysis, M.S.; 
Validation, V.A., L.B. and G.F.L.; 
Writing–Original Draft Preparation, M.S.; 
Writing–Review and Editing, V.A., L.B. and G.F.L. All authors have read and agreed to the published version of the manuscript. }	

\funding{This research received no external funding.}	

\dataavailability{All code to generate data and analysis is located at \url{https://github.com/mccsandora/Multiverse-Habitability-Handler}, accessed on. Dec. 20, 2022}

\acknowledgments{We would like to thank Cullen Blake, William Misener, Keavin Moore, and Antoine Petit for useful discussions.}
\conflictsofinterest{The authors declare no conflict of interest.}

\appendixtitles{no} 
\appendixstart
\appendix
\section[\appendixname~\thesection]{}
	
	
	Here, we collect how the physical features of our universe used throughout the text depend on the fundamental constants, displayed in Table \ref{tablefund}.
	
	\begin{table}[H]
		\caption{List of how various quantities depend on fundamental constants.  The quantities in the left column are discussed in \citep{mc1} except for $h_\text{mountain}$, which is discussed in \citep{mc3}.  The quantities in the right column were derived in \citep{mc2}.  Details of the derivations, as well as original sources where applicable, can be found within these references.  All prefactors have been chosen to reproduce values observed in our universe.}
		\label{tablefund}
		\setlength{\tabcolsep}{5mm}
		\begin{tabular}{cccc}
				
				\midrule 
				\textbf{Quantity} &\textbf{ Expression }& \textbf{Quantity} & \textbf{Expression} \\
				\midrule
				\rule{0pt}{4ex}$\rho_\text{rock}$ & $0.13 \,\alpha^3 \,m_e^3 \,m_p$ & $a_\text{ice}$ & $297.9 \,\frac{\lambda^{43/30}\,M_{pl}^{2/3}}{\alpha^4 \,m_e^{4/3} \,m_p^{1/3}}$ \\
				\rule{0pt}{4ex}$T_\text{mol}$ & $0.037\,\frac{\alpha^2\,m_e^{3/2}}{m_p^{1/2}}$ & $\Sigma(a)$ & $1.7\times10^5 \,\frac{\kappa\,\lambda^{2/3}\,M_{pl}^2}{a}$ \\
				\rule{0pt}{4ex}$M_\text{terr}$ & $92\, \frac{\alpha^{3/2} \,m_e^{3/4}\,M_{pl}^3}{m_p^{11/4}}$ & $H_\text{disk}$ & $0.2 \,\frac{\lambda^{3/80} \,m_e^{1/4} \,m_p^{1/8} \,a^{9/8}}{\alpha^{1/2}\,M_{pl}^{1/4}}$ \\
				\rule{0pt}{4ex}$R_\text{terr}$ & $3.6 \,\frac{M_{pl}}{\alpha^{1/2} \,m_e^{3/4} \,m_p^{5/4}}$ & $r_\text{disk}$ & $3.6\times10^{-6} \,\frac{\lambda^{1/3} \,M_{pl}}{\kappa\,m_p^2}$\\
				\rule{0pt}{4ex}$v_\text{esc}$ & $1.37 \,\frac{\alpha \,m_e^{3/4}}{m_p^{3/4}}$ & $t_\text{disk}$ & $4.1\times10^6 \,\frac{\lambda\,M_{pl}\,m_p^{1/4}}{\alpha^3 \,m_e^{9/4}}$ \\
				\rule{0pt}{4ex}$M_\star$ & $122.4 \,\frac{\lambda\,M_{pl}^3}{m_p^2}$ & $\omega_\text{rot}$ & $0.02 \,\frac{\alpha^{3/2} \,m_e^{3/2} \,m_p^{1/2}}{M_{pl}}$ \\	
				\rule{0pt}{4ex}$R_\star$ & $108.6 \,\frac{\lambda^{4/5}\,M_{pl}}{\alpha^2 \,m_p^2}$ & $M_\text{iso}$ & $1.3\times 10^8\,\frac{\kappa^{3/2}\,\lambda^{25/8}\,m_p^{7/4}\,M_{pl}^{9/4}}{\alpha^{15/2}\,m_e^3}$ \\
				\rule{0pt}{4ex}$L_\star$ & $9.7\times 10^{-4}\,\frac{\lambda^{7/2}\,m_e^2\,M_{pl}}{\alpha^2\,m_p}$ & $\rho_\text{halo}$ &  $2.7\times10^6\,\kappa^3\,m_p^4$\\
				\rule{0pt}{4ex}$a_\text{temp}$ & $7.6\, \frac{\lambda^{7/4}\,m_p^{1/2}\,M_{pl}^{1/2}}{\alpha^5 \,m_e^2}$ & $t_\text{migration}$ &  $10^7\,\frac{\lambda^{73/60}m_p^{1/3}M_{pl}^{5/6}}{\alpha^4m_e^{13/6}}$\\
				\rule{0pt}{4ex}$h_\text{mountain}$ &  $0.0056\,\frac{M_{pl}}{\alpha^{1/2}\,m_e^{3/4}\,m_p^{5/4}}$ & $T_\text{irr}$ & $0.19\,\frac{\lambda^{43/40}\,m_e^{1/2}\,M_{pl}^{1/2}}{\alpha\,m_p^{3/4}\,a^{3/4}}$ \\
				\midrule
			\end{tabular}
	\end{table}

\begin{adjustwidth}{-\extralength}{0cm}
\printendnotes[custom]
\reftitle{References}

\PublishersNote{}
\end{adjustwidth}
\end{document}